\documentclass[namedreferences]{kluwer}
\usepackage[dvips]{epsfig}

\begin{document}
\begin{article}
\begin{opening}

\title{\bf ENERGETICS AND DYNAMICS OF AN IMPULSIVE FLARE ON MARCH 10, 2001}

\author{RAMESH \surname{CHANDRA}$^1$, RAJMAL JAIN$^2$, WAHAB UDDIN$^1$, KEIJI YOSHIMURA$^3$, 
T. KOSUGI$^3$, T. SAKAO$^3$, ANITA JOSHI$^1$ and M. R. DESPANDEY$^2$}

\institute{$^1$Aryabhatta Research Institute of Observational Sciences, Naini Tal, 263 129, India(Formerly State Observatory) (e-mail: ramesh@aries.ernet.in)}
\institute{$^2$Physical Research Laboratory, Ahmedabad - 380 009, India}
\institute{$^3$The Institute of Space And Astronautical Science, Sagamihara 229 Japan}

\begin{ao}
Ramesh Chandra\\
Aryabhatt Research Institute of Observational Sciences\\
Manora Peak, Naini Tal$-$263 129,\\
INDIA.\\
Phone: +91-05942-235583, 235136\\
Fax: +91-05942-235136\\
email: ramesh@upso.ernet.in
\end{ao}

\begin{abstract}
We present the H$\alpha$ observations from ARIES, Nainital of a compact and 
impulsive solar flare occurred on March 10, 2001 and associated with a CME. 
We have also analysed HXT, SXT/Yohkoh observations as well as radio 
observations from Nobeyama Radio Observatory to derive the energetics and 
dynamics of this impulsive flare. We co-align the H$\alpha$, SXR, HXR, MW and 
magnetogram images within the instrumental spatial resolution limit. We 
detect a single HXR source in this flare, which is found spatially associated 
with one of the H$\alpha$ bright kernel. The unusual feature of HXR and H$\alpha$ 
sources, observed for the first time, is the rotation during the impulsive 
phase in clockwise direction. We propose that the rotation may be due to 
asymmetric progress of the magnetic reconnection site or may be due to the 
change of peak point of the electric field. In MW emission we found two 
sources, one is main source which 
is at the main flare site and another is remote source located in South-West 
direction. It appears that the remote source is formed by the impact of 
accelerated energetic electrons from the main flare site. From the spatial 
co-relation of multi-wavelength images of the different sources we conclude 
that this flare has three-legged structure.
\end{abstract}

\end{opening}

\section{Introduction}

H$\alpha$, microwave (MW), hard X-ray (HXR) and magnetogram observations are
extremely important to understand the energetics and dynamics of the solar
flares. The HXR are produced through electron ion bremsstrahlung. The MW
emission is the gyro-synchrotron radiation caused by high energy electrons
accelerated in the corona, while the H$\alpha$ comes from the thermal plasma
heated in the chromosphere by the precipitating electrons. Many studies based
on H$\alpha$, HXR and MW observations (Marsh and Hurford, 1980; Kundu, Bobrowky,
and Rust 1983; Gary and Hurford, 1990; Nishio$\it {et al.}$, 1997; Hanaoka, 1996, 2000
and references therein) have focused on whether the locations of MW and HXR sources coincide with those
of H$\alpha$ kernels.  During the impulsive phase of three flares (Marsh and Hurford, 1980)
microwave emission was dominated by a compact source located between
the H$\alpha$ kernels. In the post impulsive phase, the MW source was larger and
elongated in a direction consistent with the orientation of the magnetic field
 lines joining the H$\alpha$ kernel (Marsh and Hurford, 1980). Nishio et al.
(1997) analysed 14 impulsive solar flares to study the spatial correlation and
suggested that in the majority of impulsive flare events two loops interact
with each other, releasing magnetic energy and producing energetic electrons.
Hanaoka (1996, 2000) studied spatial correlation of different energy emissions
in many flares using H$\alpha$, Yohkoh HXT/SXT, MW and magnetogram data. In many
cases he found that the magnetic field of the flare loops show a bipolar +
remote unipolar structure, rather than a quadrapole structure.
      
    A compact and impulsive flare associated with CME occurred on 10 March
2001 during the peak of Solar cycle 23 (Uddin $\it et~al.$, 2004, hereafter referred
as paper I). This flare (1B/M6.7) occurred in NOAA AR 9368 (N27,W42) and was
well observed by ARIES, Nainital, India,  Yohkoh, SOHO missions and
Nobeyama Radio Observatory, Japan. In paper I we studied the unusual
behavior of the H$\alpha$ morphology associated with the sigmoid shaped filament
and mass motion that played a role in producing compact and impulsive flare,
which was driven by CME.
This flare was also studied earlier by Liu, Ding, and Fang (2001), St. Cyr $\it et~al.$,
(2001) and Ding $\it et~al.$, (2003). This flare is identified as an electron-rich
event in the energetic particle events of the IMP 8 data for cycle 23 (St. Cyr et
al., 2001). This flare also showed enhanced emission at continuum near the Ca
II 8542 A line. This emission lasted about 30 s, showing a good time
correlation with the peak of microwave radio flux at 7.58 GHz and thereby was
classified as type I white light flare (St. Cyr $\it et~al.$, 2001; Liu, Ding, and Fang (2001)
and Ding $et~al.$, 2003).

     In order to study the energy build-up, energy release and particle
acceleration the simultaneous multi-wavelength observations of a flare are
recently analyzed by many investigators (Gary and Hurford, 1990; White $\it et~al.$,
1992; Rolli, Wulser, and Magun, 1998; Jain $\it et~al.$, 1983; Kundu and White, 2001; Joshi,
Chandra, and Uddin, 2003). Multi-wavelength observations of bursts at millimeter
 and microwave frequencies were analysed by White $\it et~al.$, 1992 and Lin $\it et~al.$,
(1992). Based on the similarity of the time profiles of MW and HXR emission
special attention was paid (Kundu, 1961, Crannell $\it et~al.$ 1991) to understand
the flare evolution processes in these two wavebands. However, in order to
understand the principal mechanisms of energy release in these two wavebands
as well as in H$\alpha$ it is rather more important to establish the spatial
relationship of the energy sources in these wavebands. Nevertheless, this
requires high spatial resolution in both wavelength regimes (Kundu $\it et~al.$,
1989 and Nitta $\it et~al.$, 1991). On the other hand spectral information is also
crucial to determine the mechanisms that operate for the X-ray and radio
emissions. Therefore, the main purpose of this investigation is to study the
spatial correlation of the energy release sources in H$\alpha$, MW, SXR and HXR
emission by overlaying the images obtained in these wavelengths. Simultaneously
 we also study the temporal and spectral observations in these wavebands to
understand the energetics and dynamics of the flare sources.

    In preview to above goals we carried out the detailed study of the 10 March
 2001 flare. In section 2, we present the observations, while in section 3
data analysis and results are presented. We discuss our result in section 4
and conclude in section 5.

\section{Observations}

We used following observational data set for the current investigation:

1. The H$\alpha$ observations were made at Aryabhatta Research Institute of
Observational Sciences (ARIES), Nainital using 15 cm f/15 Coude telescope and
 Lyot filter centered at H$\alpha$ line. The image size was enlarged by a factor
 of 2 using Barlow lens. The H$\alpha$ images were recorded by a 16-bit 385$\times$
576 pixels CCD camera of Wright Instruments having a pixel of 22 micron$^2$.
The resolution of the image is 1 arcsec per pixel.

2. Soft X-ray data from SXT onboard Yohkoh (Tsuneta $\it et~al.$, 1991). The SXT
provides soft X-ray images with a resolution of 2.5 arcsec per pixel and temporal 
resolution of 2 sec during the flare.

3. Hard X-ray data from HXT onboard Yohkoh, which is a Fourier synthesis type
telescope (Kosugi $\it et~al.$, 1991). HXT provides observations in four energy bands
i.e. L (14-23 keV), M1(23-33 keV), M2 (33-53 keV) and H (53-93 keV). Its
angular and temporal resolution are 5 arcsec and 0.5 sec respectively during
the flare.

4. Magnetograms from full disk Michelson Doppler Imager (MDI) onboard SOHO
mission. MDI takes full disk magnetogram on a 1024 X 1024 CCD array with a
spatial resolution of 2 arcsec per pixel (Scherrer $\it et~al.$, 1995), and temporal
resolution of 90 minute.

5. The radio data from Nobeyama Radioheliograph (Nakajima $\it et~al.$, 1994; Takano
$\it et~al.$, 1997). The Radioheliograph has angular resolution of 10 arcsec and 5
arcsec at 17 and 34 GHz respectively.

To understand the spatial relationship between different sources in this investigation
we co-align the SXR, HXR and MW data with the H$\alpha$ data. The co-alignment
of SXR, HXR and MW data were achieved by referring to relative position from the
solar disk center of each field of view. As for the H$\alpha$ observations, we needed
to get such a position information through the comparison between them and white light
images from SOHO/MDI of same observation time. The accuracy of this co-alignment are 
better than 5 arcsec.

   All the data analysis has been done by the Solarsoft and IRAF software
tools.

\section{Analysis and Results}

\subsection{H$\alpha$ OBSERVATIONS}

   In Figure 1 we show the sequence of H$\alpha$ filtergrams that indicate the
evolution of the flare and bright and dark mass ejections associated with the
flare activity. We noticed considerable pre-flare activity in the filament
such as twisting motion and perhaps rising up as shown earlier by Singh and
Gupta (1995) and recently by Sterling et al (2001), pre flare plage
brightening as discussed earlier by Jain (1983) and Priest (1984). The
pre-flare brightening started around 03:39:35 UT. 
  
   The disruption of the H$\alpha$ filament channel, which is connected to the sigmoid 
filament was noticed between 03:54:06 - 03:54:51 UT in the following part
of the active region. The filament channels are the primary consistents of a 
chromospheric coronal environment in which the filaments are supported by magnetic fields 
and a single filament channel may survive successive eruptions and reformation of 
its filament (Gaizauskas et al. 1997; Gaizauskas, Mackay, and Harvey, 2001). 

    The MDI magnetograms and H$\alpha$ observations of the active region showed that 
the positive parasitic polarity emerged near the following sunspot, where the flare
occurred (paper I). The flare started around 04:01:11 UT as a two small
bright kernels  K1 and K2 (cf. Figure 1). The intensity of both kernels
rapidly increased. However, the kernels were confined in small area and the
flare was compact but impulsive as discussed in paper I.

   Figure 2 illustrate the evolution of flare during the impulsive phase. Inspection 
of these H$\alpha$ contours of the flare we noticed that the H$\alpha$ source is 
rotating in the clockwise direction as we have seen in HXR data (c.f. subsection 3.3). 
The rotation in the source is very interesting feature of this flare.

   During the peak phase around 04:04:16 UT bright material ejected, which
we call as bright mass ejecta (BME). We estimated the initial velocity of BME
600 km sec$^{-1}$ when it started from North-West direction of the flare location.
Around 04:06:27 UT the BME was followed by dark mass ejecta (DME), which,
however, was moving with a slower speed of about 400 km sec$^{-1}$ in the early
stage and then speed decayed exponentially very fast. Apparently, the whole
flare material erupted away as bright and dark mass ejections. Almost after
the decay phase of the flare some dark matter was seen in the form of dark
loops like filamentary system, which we call Filamentry Dark Material (FDM) 
in North-East direction of the flare site.
This FDM was perhaps formed due to condensation of the ejected
material having speed less than escape velocity. The detailed discussions
about BME, DME and their association with CME have been given in paper I.

\begin{figure}[t]
\vspace*{-0.2cm}
\hspace*{-0.1cm}
\epsfig{file=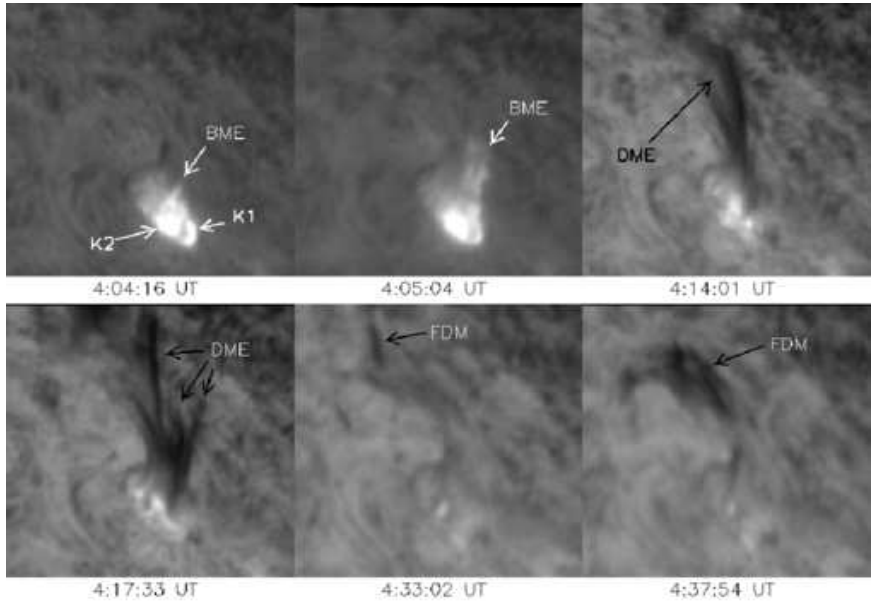,width=11.5cm,height=8cm}
\vspace*{-0.1cm}
\caption{H$\alpha$ filtergrams of the flare. The BME, DME and the condensation of filament 
dark material (FDM) are shown by arrows. The FOV of the images is 160$^{''}$ $\times$ 145$^{''}$. The North is up and East is to the left.}
\end{figure}

\begin{figure}
\vspace*{-9cm}
\hspace*{-1cm}
\epsfig{file=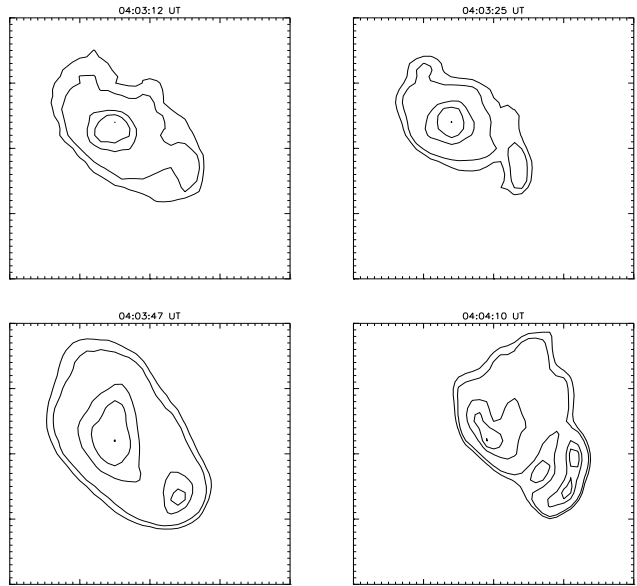,width=15cm,height=20cm}
\vspace*{-4cm}
\caption{H$\alpha$ contours of the flare during the impulsive phase.
The contour levels are the 55 $\%, 60 \%, 70 \%, 80 \%, 90 \% and 100\%$ of 
the peak counts. The FOV is 40$^{''}$ $\times$ 40$^{''}$. The North is up 
and East is to the left.}
\end{figure}

\subsection {SPATIAL CORRELATION OF H$\alpha$, SXR, HXR AND MW SOURCES} 

In this section we present H$\alpha$ images overlaid by contours of SXT/AlMg
images, HXT images (L and H band), 17 GHz MW images (I and V) and 34 GHz at
different phases of the flare to study the spatial correlation of the energy
sources that emitting at different wavebands.

\begin{figure}[h]
\hbox{
\psfig{file=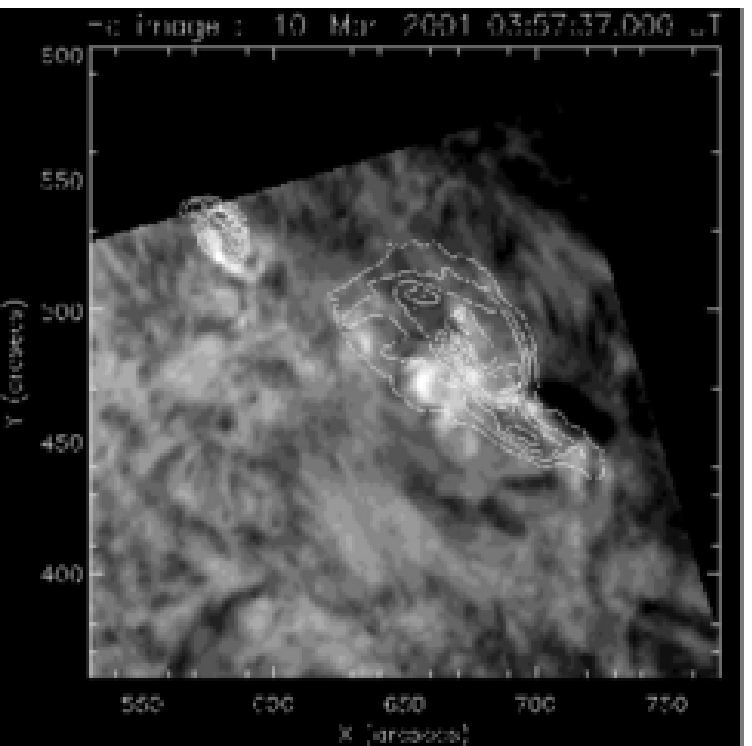,height=6.0cm,width=6.0cm}
\hspace*{-0.5cm}
\psfig{file=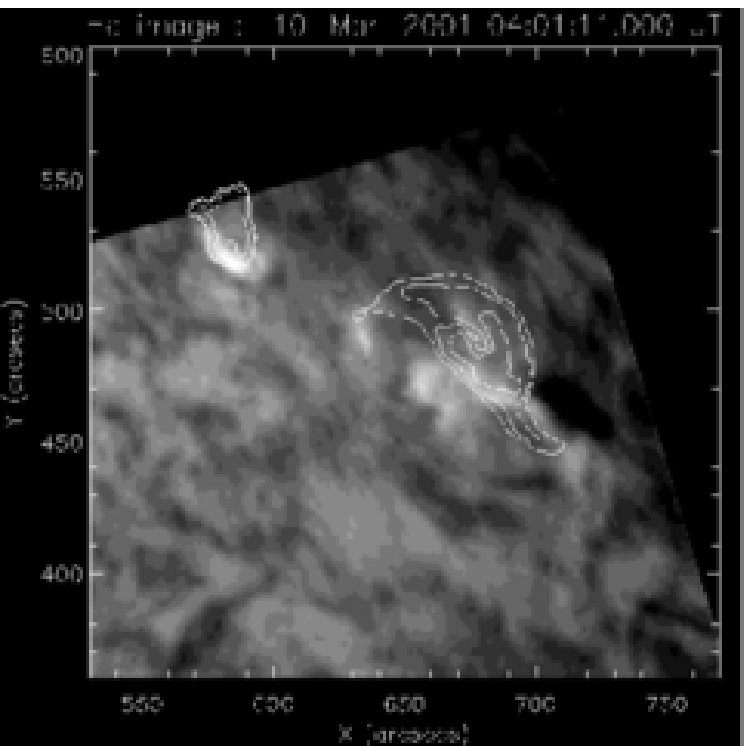,height=6.0cm,width=6.0cm}
}
\caption{H$\alpha$ images (gray scale) overlaid by SXR contours. 
North is up and East is to the left.}
\end{figure}

\begin{figure}[h]
\hbox{
\psfig{file=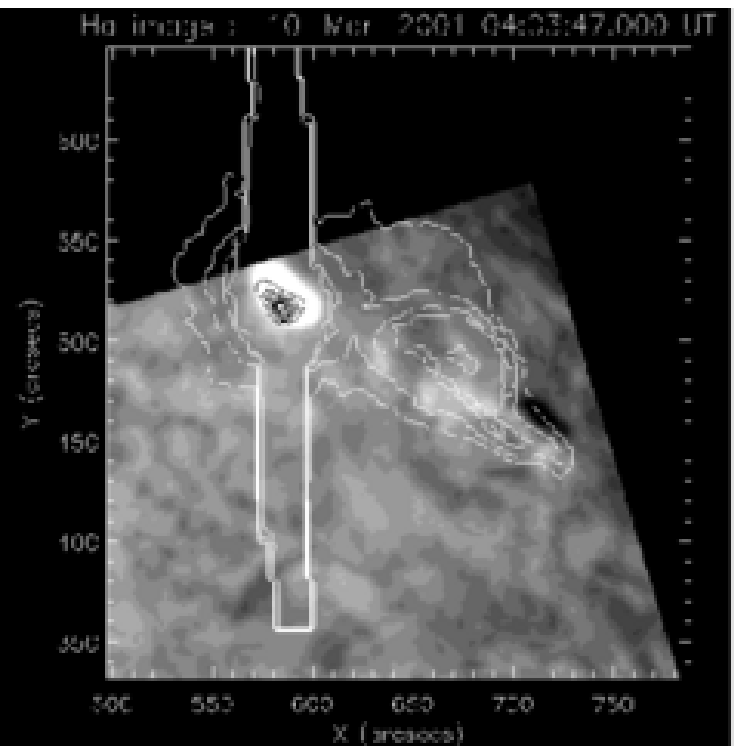,height=6.0cm,width=6.2cm}
\hspace*{-0.8cm}
\psfig{file=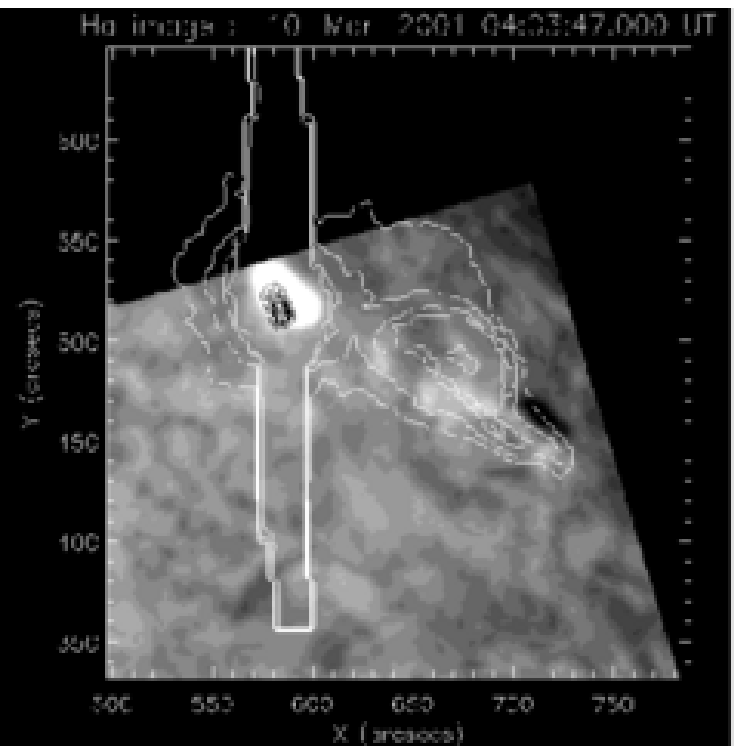,height=6.0cm,width=6.2cm}
}
\caption{H$\alpha$ images (gray scale) overlaid by SXR (white contours) 
and HXR  L (left) and H (right) band (black contours). Contour level 
are 25, 35.4, 50 and 70.7 $\%$ of the peak count. North is up and East 
is to the left.}

\end{figure}

Figure 3 shows SXT/AlMg images (contours) overlaid on the H$\alpha$ images
(grey scale) within instrumental spatial resolution limit at 03:57:37 UT
(about 3 minute before the flare onset) and at 04:01:11 UT (near the flare
onset) respectively. In the precursor phase there was plage brightening near
the leading spot group as well as at the flare location. It may be noted that
the bright SXR sources (contours) were also present on both locations before
flare (cf, Figure 3, left) in consistent to H$\alpha$ morphology described in
former section 3.1 and discovered earlier by Jain (1983). The H$\alpha$ image
taken at the onset of the flare at 04:01:11 UT (cf, Figure 3, right) showed
enhanced emission in H$\alpha$ as well as in SXR (contours) in the following
spots of the active region. This overlaid plot suggests that SXR brightening
is above the H$\alpha$ brightening in the coronal loops and they are very
extended sources while H$\alpha$ sources are very compact as observed by us.

\begin{figure}
\vspace*{-3cm}
\hspace*{-1.0cm}
\epsfig{file=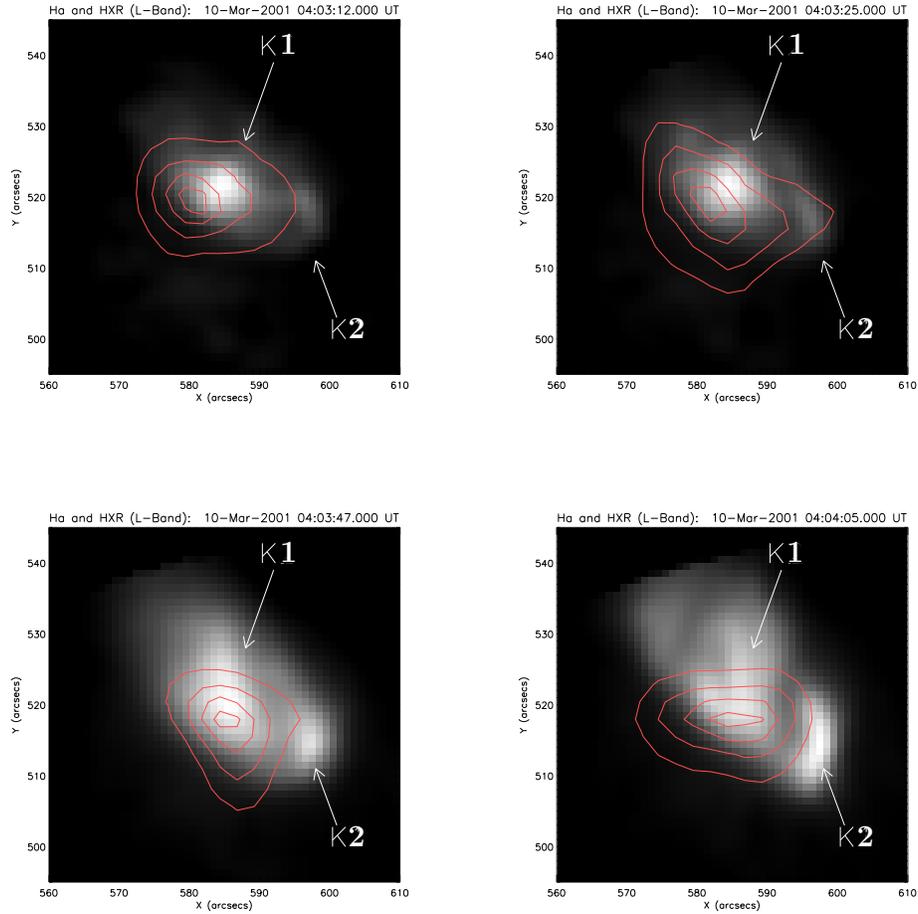,width=14cm,height=20cm}
\vspace*{-4cm}
\caption{H$\alpha$ images (gray scale) overlaid by  L-band HXR contours.
Contour level are 25, 35.4, 50 and 70.7 $\%$ of the peak count. North is up and East is to the left.}
\end{figure}

\begin{figure}
\vspace*{-3cm}
\hspace*{-1.0cm}
\epsfig{file=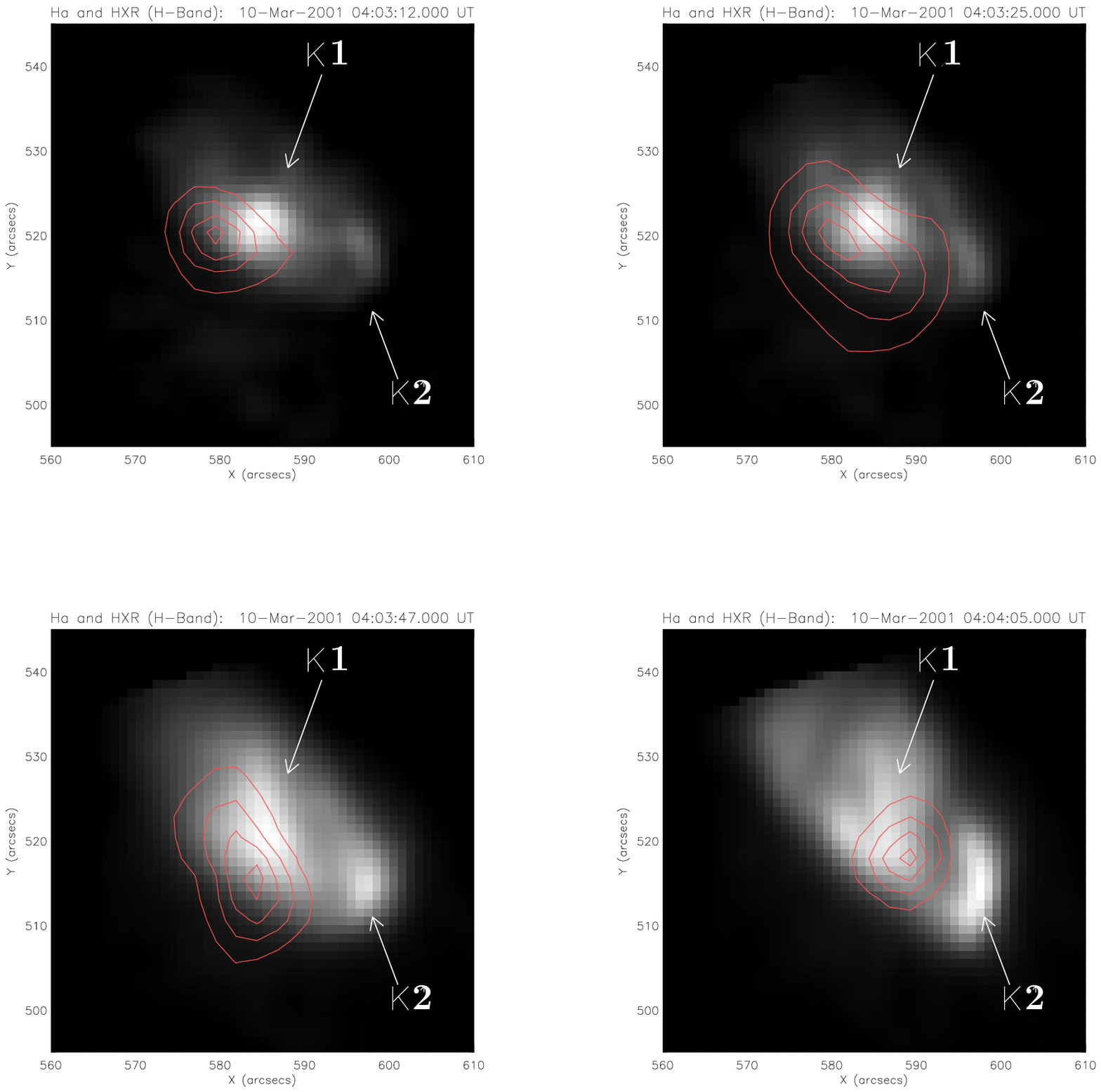,width=14cm,height=20cm}
\vspace*{-4cm}
\caption{H$\alpha$ images (gray scale) overlaid by H-band HXR contours.
Contour level are 25, 35.4, 50 and 70.7 $\%$ of the peak count. North is up and East is to the left.}
\end{figure}

We show in Figure 4 (left and right) overplots of H$\alpha$, SXR (white
contours) and HXR (L and H band - black contours) at 04:03:47 UT during
impulsive phase of the flare. This Figure shows unambiguously that HXR
sources (both L and H bands) are compact with respect to SXR source and
their positions coincide with the H$\alpha$ flare emission sources (kernels K1
and K2).

\begin{figure}[h]
\vspace*{-2cm}
\hspace*{-1.5cm}
\epsfig{file=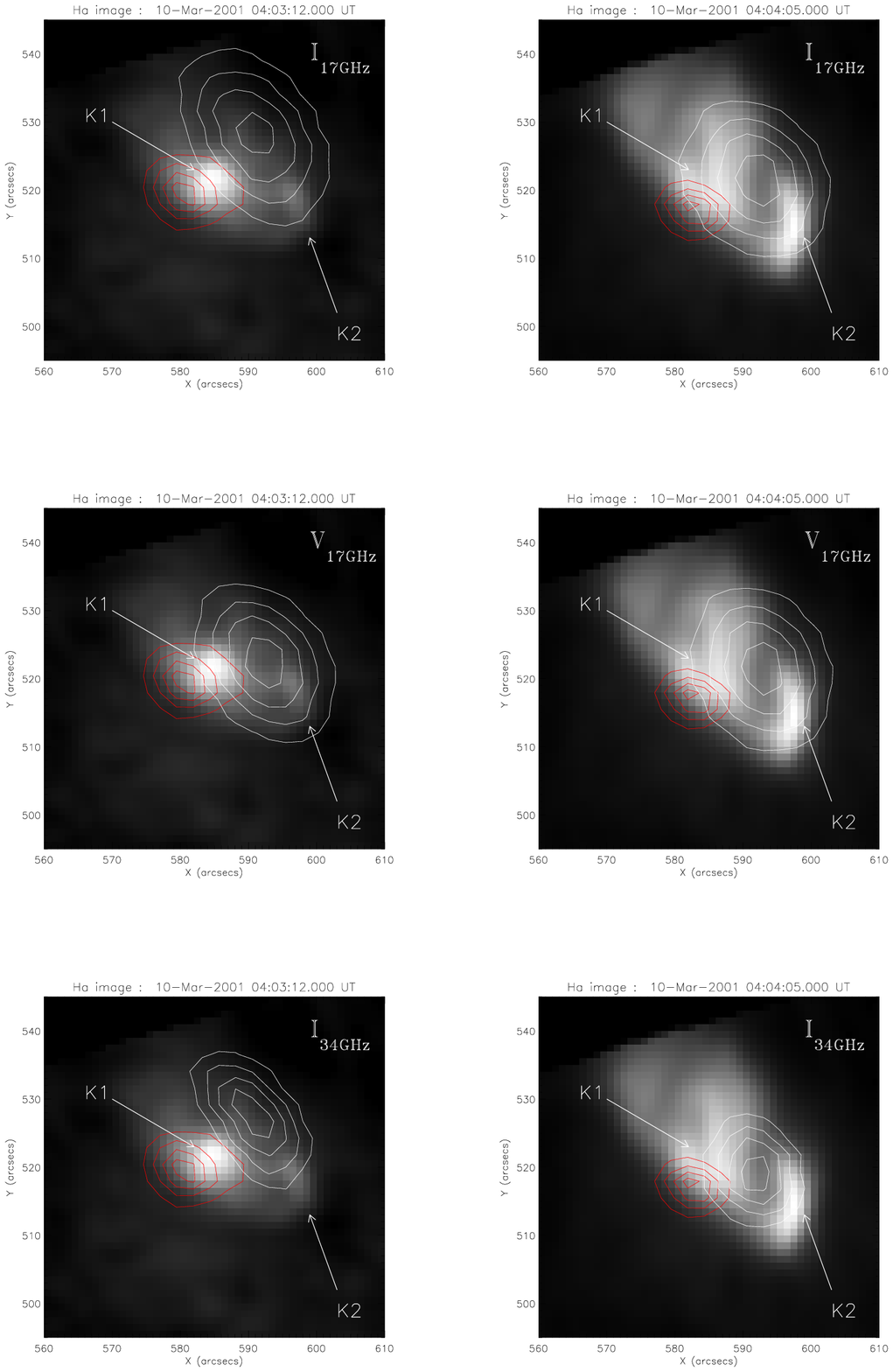,width=14cm,height=20cm}
\vspace*{-3.5cm}

\caption{H$\alpha$ images (gray scale) overlaid by HXT L-band (red contours, level=25, 35.4, 50 and 70.7 $\%$ of 
the peak T$_b$ ) and MW ( 17 GHz I, V and 34 GHz I) (white contours, level=50, 60, 75 and 90 $\%$ of the peak T$_b$). 
North is up and East is to the left.}

\end{figure}

In order to understand better spatial correlationship of H$\alpha$, HXR and
 MW sources we selected smaller area 50$\times$50 arcsec squire of the flaring
region to overlaid HXT (L and H) and MW (I and V) images on H$\alpha$ flare
images. These overlaid images showed spatial correlation more precisely.
The overlaid images (contours) of HXT L and H band on H$\alpha$ image (gray
scale) have been shown in Figure 5, 6 and 7. The overlaid HXT L and H band
images on H$\alpha$ at 04:03:12 UT, 04:03:25 UT, 04:03:47 UT and 04:04:05 UT
showed temporal evolution of the spatially correlated energy release sources
in these two wavebands.

\begin{figure}[t]
\vspace*{-6cm}
\hspace*{-2cm}
\epsfig{file=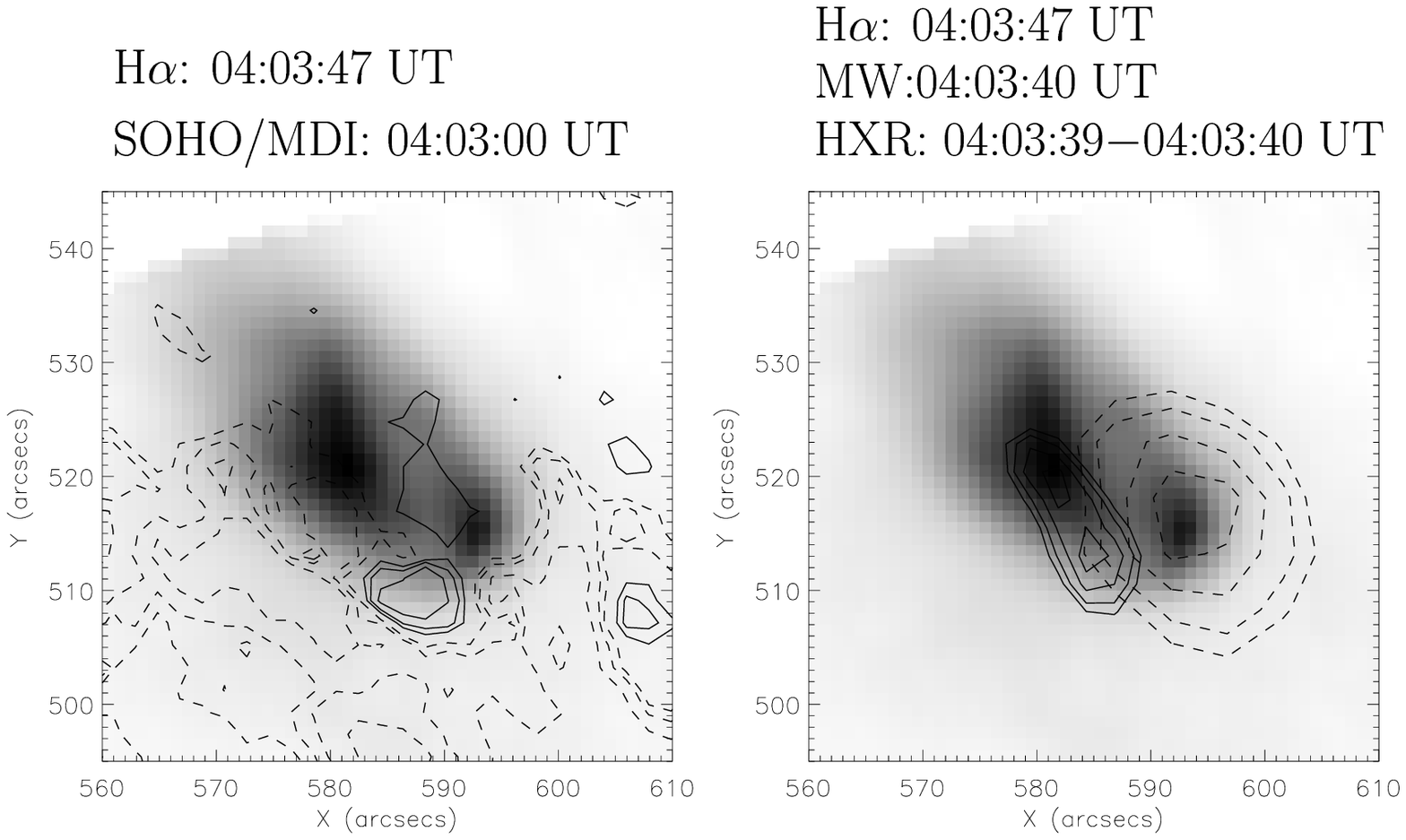,width=15cm,height=20cm}
\vspace*{-7cm}
\caption{The H$\alpha$ negative image overlaid by magnetic field contours ( solid line N
polarity and dotted line S polarity, level=$\pm 50, \pm 100,\pm 200,\pm 500 $ G)(left) 
and H$\alpha$ image overlaid by HXT H- band (solid contour, level=50, 60, 75 and 90 $\%$ 
of the peak counts) and MW 17 GHz. (dotted contour, level= 50, 60, 75, 90 $\%$ of peak T$_b$) (right).}
\end{figure}

    In Figure 7 at 04:03:12 and 04:04:05 UT we have plotted HXT L band image
(red contour) and 17 GHz MW I and V sources (white contours) on the H$\alpha$
images (gray scales). In this Figure we also present the overlaid images of
H$\alpha$ (gray scales), HXR L band image (red contour) and 34 GHz I image
(white contours). In 34 GHz we have better spatial resolution than 17 GHz.
The 17 GHz (V) sources in middle panel of Figure 7 were almost
located at the same positions. This indicate that there is no significant
change in the coronal magnetic polarities.
 
     Our above co-alignment study revealed that compact single HXR source was
laid above the H$\alpha$ K1 kernel, while the MW source was between the kernel
K1 and K2. These sources were very close to each other.
The estimated distances between center to center of HXT L band and 17 GHz I
sources was 15 arcsec and on the other hand HXT L band and 17 GHz V was 12
arcsec. The distance between the 34 GHz and HXR L band sources was
15 arcsec. The position difference between the MW and HXR sources are may be due 
to the possibility of projection effects, since different emission sources may
come from different height of solar atmosphere. Thus our analysis showed that HXR 
and MW sources are almost co-align to 
each other (cf. Figure 7). On the other hand the HXR source was very close 
to H$\alpha$ kernel K1. The above scenario indicates that the energy release site 
was in the low lying loops and therefore the emission originated from the lower 
lying loops.

    Figure 8 (left) represents the overlaid images of MDI magnetograms contours
(solid line N polarity and dotted line S polarity) on H$\alpha$ (gray scale).
On the other hand, in the right panel we show H$\alpha$ image overlaid by image
of HXR H band (solid line) and MW 17 GHz (dotted contour). 
This Figure revealed that this flare has very compact sources
in H$\alpha$, HXR and MW emission and they are almost co-aligned. From this
Figure it is further clear that magnetic field at the flare location is very
complex, viz. the emerging flux of N polarity penetrated into the S polarity,
 which might have triggered the flare.

\begin{figure}[h]
\vspace*{-3.0cm}
\hspace*{-2.8cm}
\epsfig{file=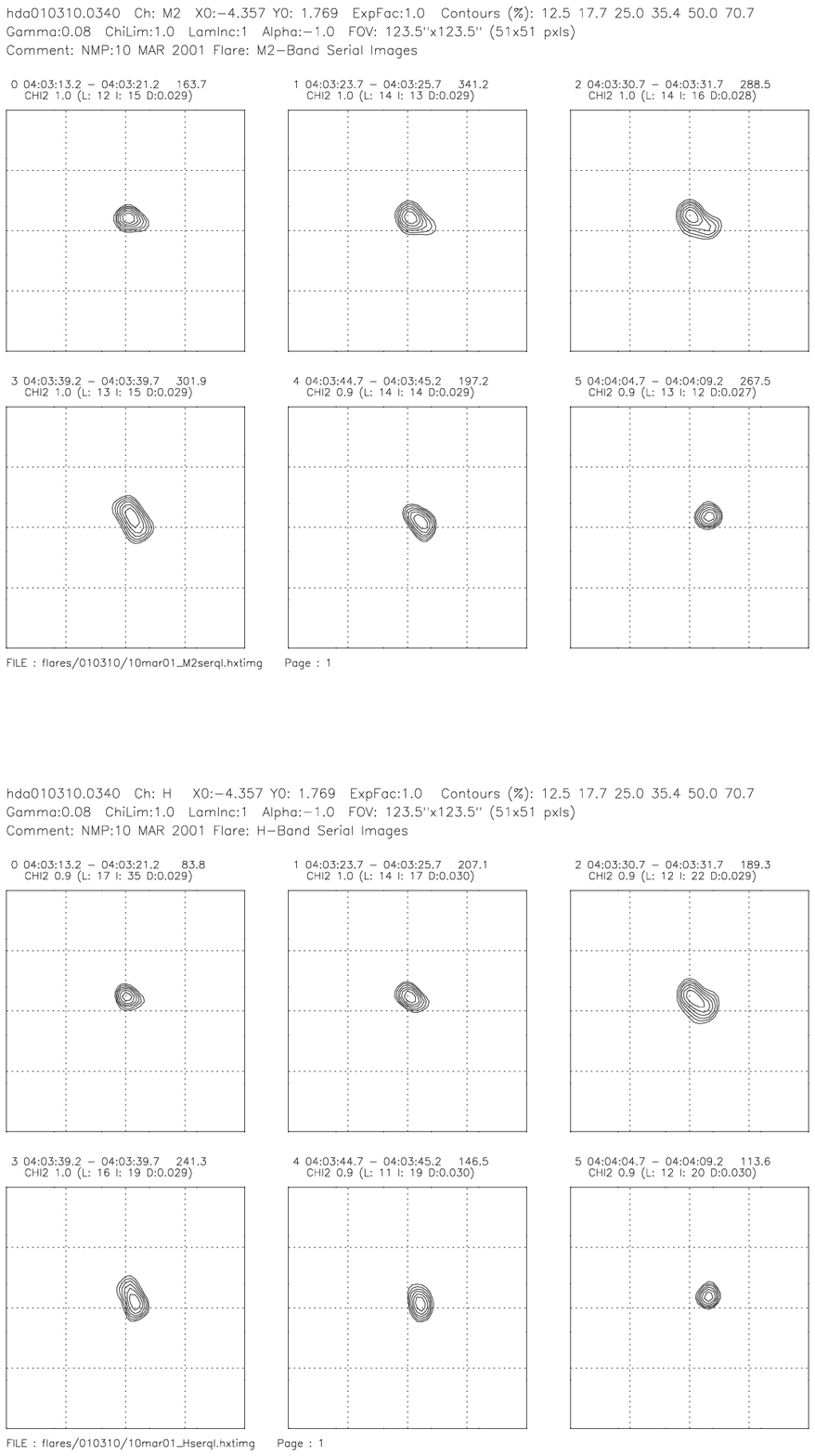,width=18cm,height=24cm}
\vspace*{-5.8cm}
\caption{ HXR M2 and H band images (contour) the flare during the impulsive phase.}
\end{figure}

Around 04:03:51 UT a remote brightening (c.f. Figure 12) in MW emission
appeared in South-West direction i.e. towards the leading spot, which, however,
 showed right handed polarization. This implies that the main and the remote MW
 emission have their source in opposite magnetic polarities. It also seems that
 this remote source occurred by the impact of accelerated energetic electrons
from the reconnection site.

\subsection{DYNAMICS OF HARD X-RAY SOURCES}

Yohkoh/HXT observed this flare in great detail. The four HXT energy channels
L, M1, M2 and H showed strong impulsive burst with almost equal counts 528,
570, 550 and 463 cts/sec/sc respectively, indicating very hard spectrum. The
impulsiveness of the flare following Pearson $\it et~al.$, (1989) and Jain $\it et~al.$,
(2002) in hard X-ray is 6.06 counts sec$^{-2}$.

\begin{figure}[h]
\hspace*{0.5cm}
\epsfig{file=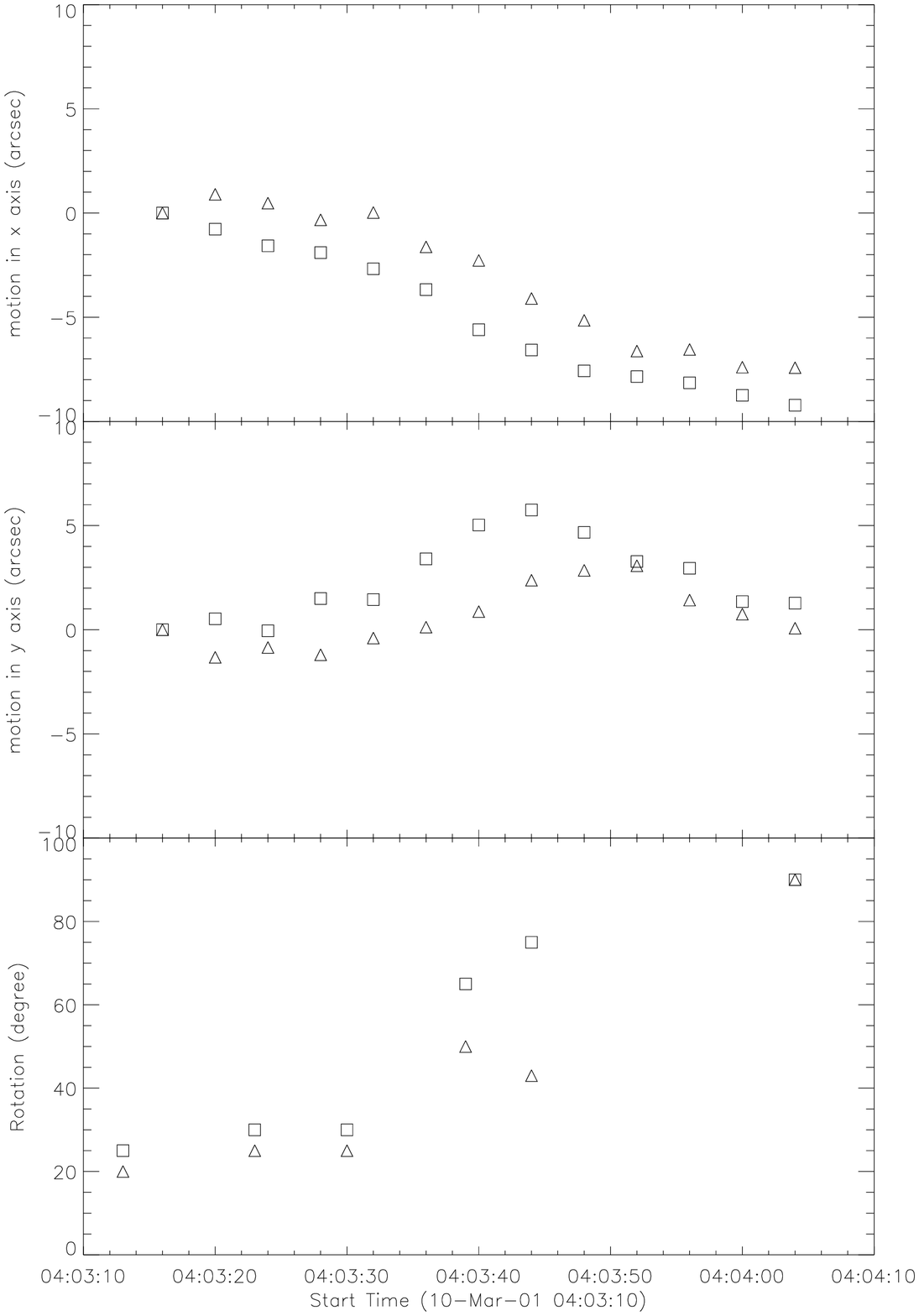,width=10cm,height=14cm}
\vspace*{0.5cm}
\caption{ Temporal variation of the central position of HXR source in x (West) , y (North)
direction taking the 04:03:16 UT image as reference and HXR source rotation from x-direction 
in clockwise direction. The square and triangle represents the HXR M2 and H energy band respectively}
\end{figure}

In Figure 9 we show the temporal evolution of HXR image (contour) in the
M2 and H bands during the impulsive phase of the flare. In all energy bands
only single source was seen. We have calculated the size of HXR source using
the ellipse fit on source. From this ellipse fit the estimated size of the
HXR source was 7.0$\times$3.8, 6.7$\times$3.3, 6.5$\times$3.3 and 6.2$\times$3.2 
arcsec in L, M1, M2 and H energy bands respectively.
   
   Figure 10 shows that the HXR source moves in North-West direction and rotates
in clockwise direction. The motion of HXR source in x (West) direction was
found more than y (north) direction. To study the motion as well as rotation
related to HXR source in Figure 10 we plot the temporal variation of central
position of the source and the angle of source from the x-axes in clockwise
direction. To determine the centroid location of HXR source, a two-dimensional
noncircular gaussian fit was used. The motion in HXR sources is common, which
 has been studied by several authors (Sakao, 1992; Krucker, Hurford, and Lin, 2003;
Bogachev $\it et~al.$, 2005 and references therein). 
The rotation of the HXR source in this impulsive flare appears an unusual phenomena
 and being reported by us for the first time. We have measured average speed
and rotation rate of HXR source which is about 100 km sec$^{-1}$ and 1.5 deg sec$^{-1}$
respectively.

  In Figure 11 we show the X-ray photon spectrum derived by the Yohkoh/HXT
data at the maximum of the hard X-ray flare. It may be noted that the spectrum
is well fitted by power-law with power law index 2.4 which indicate the very
hard spectrum. This result suggests that the HXR emission above 14 keV was
produced by non-thermal electron beams. In table I we present the parameters
derived from the single power-law fit as shown in Figure 10. In order to derive
the plasma parameters we used the following formula:

\begin{equation}
I(E) = A_{1} E^{-\gamma} photons \times cm^{-2} s^{-1} keV^{-1}
\end{equation}

\noindent Where I is the photon flux at an energy E in keV, $\gamma$ is power-law index and A$_{1}$
is a constant. According to thick target model the total energy flux of non-thermal
electrons above the cut-off energy E$_{0}$ is given by the following formula (Croshya, Aschwanden, and
Dennis, 1993 and Tomczak, 1999):

\begin{equation} 
E^\prime (\geq E_{0}) = 4.8 \times 10^{24} A_{1} E_{0}^{-\gamma+1} \gamma(\gamma-1) B(\gamma - 0.5, 0.5) erg \times s^{-1}
\end{equation} 

\noindent Where B is the beta function. In the considered case the calculated total 
energy flux of non-thermal electrons having energies above 14 keV was
1.1 $\times 10^{29}$ ergs sec$^{-1}$.

\begin{figure}

\hspace*{-1cm}
\epsfig{file=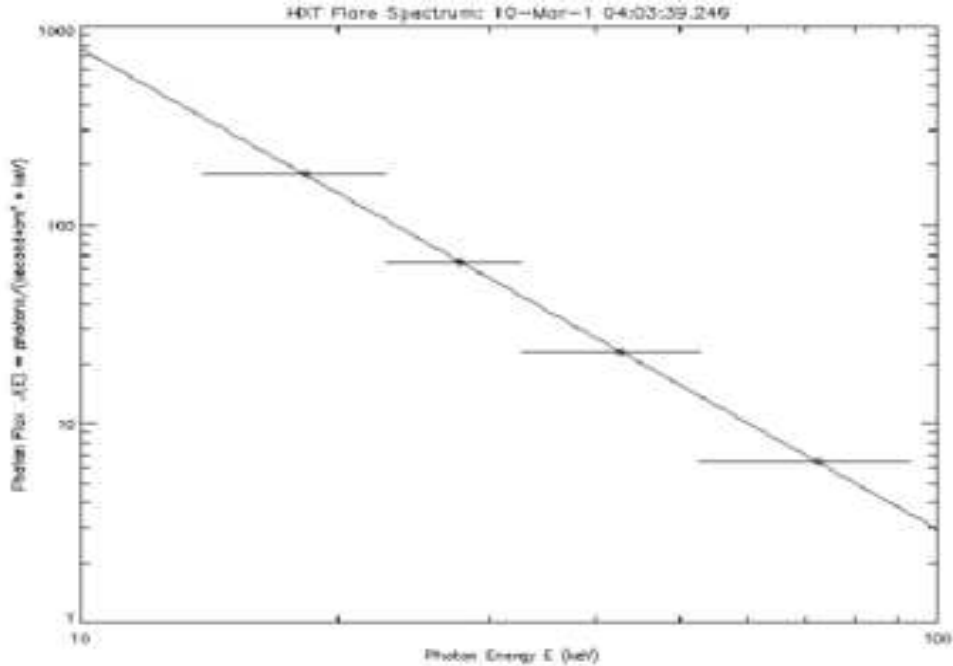,width=13cm,height=9cm}
\vspace*{0.5cm}
\caption{ HXR Spectra derived from HXT/Yohkoh data at the maximum of the flare. }
\end{figure}

\begin{table}[h]
\caption{Parameters derived from HXT data}
\medskip
\label{tab:specs1}
\begin{tabular}{ll}
\hline

Power-law Index, $\gamma$ & 2.4 \\
Parameter A$_{1}$ in Eq. 1 & 2.02 $\times$ 10$^5$ \\
Total energy flux of non-thermal electrons, &\\
E$^\prime$(E$\geq$ 14keV)[ergs s$^{-1}$] & 1.1 $\times$ 10$^{29}$ \\
Duration of impulsive onset, $\Delta$t [s] & 20\\
Total energy deposition by the E $\geq$ 14 keV & \\
electrons, (1/2)E$^\prime$(E$\geq$ 14 keV) $\times$ $\Delta$t [ergs] & 1.1 $\times$ 10$^{30}$\\

\hline
\end{tabular}
\end{table}

\section{Discussion}

From the study of multiwavelength observations of impulsive flare of
March 10, 2001 made in H$\alpha$, SXR, HXR and MW we found that this flare was
composed of compact sources in H$\alpha$, SXR, HXR and MW emission.

\begin{figure}[t]
\vspace*{-10cm}
\hspace*{-0.3cm}
\epsfig{file=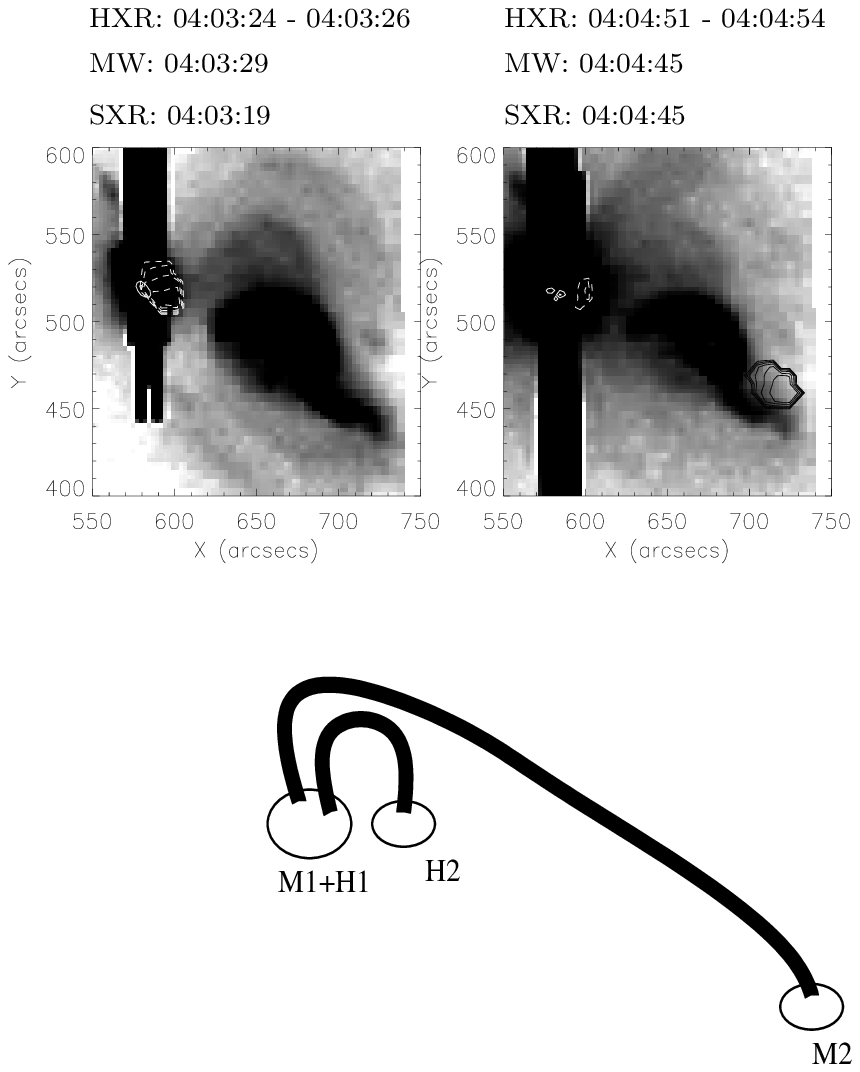}
\vspace*{-9cm}
\caption{Coaligned SXR images overlaid by HXT M1 band (white contour, level = 50, 90 $\%$of the 
peak intensity and MW 17 GHz (V) contour (black solid (+ve) and white dashed line (-ve),
level = $\pm 3, \pm 5, \pm 7, \pm 10, \pm$ 20 $\%$ of peak counts) (upper panel). Lower panel
is cartoon representing the spectra relationship between MW, HXR and SXR sources. The MW sources
are shown by M1, M2, while unresolved HXR sources are shown by H1 , H2.}
\end{figure}
   
From the co-alignment of H$\alpha$, HXR and MW source it is clear
that the HXR source seems to be located near the footpoint on kernel K1. The
above topology of the HXR, MW and H$\alpha$ emission suggests that the cause
for these emissions is energetic electrons. It appears that the emerging flux
region (cf. paper I), close to following sunspot, was showing considerable
activity and was approaching towards the following sunspot. This caused a
disruption of filament channel, which apparently opened up the pre-existing
loop system and led the reconnection between low lying loops of EFR and high
lying loops of main active region. During this reconnection we feel that
electrons were accelerated to very high energies and propagated very fast
towards the feet of the low lying loops causing HXR emission as well as
H$\alpha$ emission. It was also noticed by us and reported in paper I that
magnetic field was highly strong and sheared, which suggests that in addition
to non-thermal bremsstrahlung also gyro-synchrotron process might be operating
causing to MW emission as observed by us associated with this flare. However,
from the similar time profile and evolution of HXR and MW emission it seems
that these two emissions were produced from the same population of non-thermal
electrons and they transferred their energy to the chromospheric material by
collision while passing through the chromosphere (Fisher, Canfield, and McClymont, 1985; Brown,
1973 and Jain $\it et~al.$, 2000, 2005).

    The HXR emission was from a single and compact source in all energy bands.
According to Sakao et al. 1992 the single HXR source is produced if asymmetry of 
fluxes from double sources is so large, due to asymmetry in magnetic field strength 
around the double sources,
that the flux from one of the pair of double source is below the dynamic range
of Yohkoh/HXT, then the other source would be observed as a single source. In
the present flare case, which occurred at N27W42 location it does not appears
to us that strong magnetic field asymmetry might be existing in the region.
However, Sakao et al. (1992) also suggested some other possibilities such as
(1) the separation of double sources is smaller than the spatial resolution
of HXT, hence an apparent single source is imaged, and (2) HXR emitted
near the top of the flaring loop. Thus, within the spatial resolution limit,
we propose that the observed HXR single source may be a double source but it
was not resolved.

     In HXR source we noticed the motion in the North-West direction. The HXR
 source motion has been studied earlier by many authors (Krucker, Hurford, and Lin, 2003;
Ding $\it et al.$, 2003; Sakao $\it et~al.$, 1994; Sakao $\it et~al.$, 2000 and references therein).
The source motion in M2 and H energy bands was found almost in the
same direction. We interpret the HXR source motion as the shifting of the foot
 points as a function of the reconnection.
The HXR source at a foot point is generated as a result of
interaction of accelerated electrons with the ambient material near foot point
of the loop and the source is progressively changing its location as the newly
magnetic field lines get reconnected.
     
    In this impulsive flare we also noticed the rotation in HXR and H$\alpha$
source in clockwise direction, which is an unusual feature. To explain this rotation we
assume that the pair of two unresolved HXR sources corresponds to the two ends
of a series of reconnected magnetic loops. In this situation, one possibility
is that the shear angle is different for low lying and high lying loops and
the shear angle is changing. Due to this change of shear angle the site of
magnetic reconnection will also change and hence the HXR source is rotating.
The other possibility of HXR source rotation is that two unresolved HXR foot
points may move same direction along the neutral line with different speeds. 
Such motions are the chromospheric signature of displacement of the particle 
acceleration region during the impulsive phase ( Bogachev $\it et~al.$, 2005).
The observations by Wang $\it et~al.$ (2003) indicate that an electric field in the corona
is not uniform along the reconnecting current layer at the separator. The peak
point of the electric field may change its position during the impulsive phase
of the flare. Consequently the unresolved HXR sources should move in the same 
direction along the neutral line and hence the HXR source may be rotating.
    
    From the spatial correlation between different waveband sources we
conclude that the analysed flare had the 'three-legged' structure i.e. it may
be considered to be one of the typical configuration of loops as suggested
earlier by Hanaoka (1996). Accordingly it might be possible that the disruption
 of filament channel and the flare were caused as a consequence of interaction
between low lying loops of EFR and long high lying loops of pre-existing main
active region. Our results suggest such mechanism in consistent to earlier
findings of Hanaoka (1996, 1997) and Nishio $\it et~al.$ (1997). 

       The observations of the flare under study and results obtained from
their analysis reveal a scenario as shown in Figure 12.  The MW source at the
flare site is marked as M1 main source and remote MW source is marked as M2
and unresolved single HXR source is marked as H1 and H2 at the main flare site,
 which is coinciding with H$\alpha$ kernel K1. This cartoon is similar to that
of Hanaoka (1996, 1997).

\section{Conclusion}

We have studied the co-alignment of H$\alpha$, SXR, HXR, MW and Magnetograms of 10 March, 2001 impulsive flare and the main results are as follows:
                                                                                                          
1. The HXR source was found spatially associated with H$\alpha$ bright kernel K1 and MW source seems to be on the loop top.

2. The analysed impulsive flare possibly had three-legged structure.

3. We detect a single HXR source, which showed motion in North-West direction as well as 
rotation in clockwise direction. 
We also noticed the clockwise rotation in H$\alpha$.
The rotation in the HXR and H$\alpha$ sources are an unusual feature 
of this flare. To explain this rotation we propose the following two 
possibilities:

(a) The rotation may be result of progressive reconnection of 
magnetic field lines and thereby continuously changing the magnetic reconnection site.

(b) The peak point of the electric field may change during the impulsive phase, hence
the HXR and H$\alpha$  sources are rotating.

\begin{acknowledgements}

\noindent This work was supported by Japan Society for Promotion of Science (JSPS) and Department
of Science and Technology (DST), India under India-Japan Cooperative Science Programme (IJCSP).
We are thankful to Prof. Satoshi Masuda, Prof. B. V. Somov, Prof. K. Shibasaki and Prof. T. Sakurai
for very useful discussions which help in the interpretations of our results. The authors
(R.C. and W.U.) are also thankful to Prof. Ram Sagar for his kind support to carry out
this work. Special thanks to the anonymous referee for his/her constructive comments 
and suggestions which improved the scientific content of the paper significantly.

\end{acknowledgements}

\end{article}

\end{document}